\documentclass{iau} 
\usepackage{graphicx}
\usepackage{color}
\usepackage{multicol}

\title[PIC: Global Relativistic Jets with Helical Magnetic Fields ] %% give here short title %%
{Particle-in-cell Simulations of \\ Global Relativistic Jets with Helical Magnetic Fields}

\author[Ioana Du\c{t}an et al.]   %% give here short author list %%
{Ioana Du\c{t}an$^1$, Ken-Ichi Nishikawa$^2$, Yosuke Mizuno$^3$, Jacek Niemiec$^4$, Oleh Kobzar$^4$, Martin Pohl$^{5,6}$, Jose L. G\'{o}mez$^7$, Asaf Pe'er$^8$,\\ Jacob T. Frederiksen$^9$, \r{A}ke Nordlund$^9$, Athina Meli$^{10}$, Helene Sol$^{11}$, \\Philip E. Hardee$^{12}$, and Dieter H. Hartmann$^{13}$}

\affiliation{
$^1$Institute of Space Science, Atomi\c{s}tilor 409, Bucharest-Magurele RO-077125, Romania\\ email: {\tt idutan@spacescience.ro}\\ [\affilskip]
$^2$Department of Physics, University of Alabama, Huntsville, AL 35899, USA\\ email: {\tt ken-ichi.nishikawa@nasa.gov} \\ [\affilskip]
$^3$Institute for Theoretical Physics, Goethe University,  Frankfurt am Main {D-60438}, Germany\\ email: {\tt mizuno@th.physik.uni-frankfurt.de} \\ [\affilskip]
$^4$Institute of Nuclear Physics PAN, ul. Radzikowskiego 152,  Krak\'{o}w 31-342, Poland\\ email: {\tt Jacek.Niemiec@ifj.edu.pl (J.N.); oleh.kobzar@ifj.edu.pl (O.K.)} \\ [\affilskip]
$^5$Institut fur Physik und Astronomie, Universit\"{a}t Potsdam,  Potsdam-Golm {D-14476}, Germany\\ [\affilskip]
$^6$DESY, Platanenallee 6,  Zeuthen {15738}, Germany\\ email: {\tt marpohl@uni-potsdam.de} \\ [\affilskip]
$^7$Instituto de Astrof\'{i}sica de Andaluc\'{i}a, CSIC, Apartado 3004, Granada 18080, Spain\\ email: {\tt jlgomez@iaa.csic.es} \\ [\affilskip]
$^8$Physics Department, University College Cork, Cork T12 YN60, Ireland\\ email: {\tt a.peer@ucc.ie} \\ [\affilskip]
$^9$Niels Bohr Institute, University of Copenhagen, Blegdamsvej 17, Copenhagen DK-2100, Denmark; email: {\tt trier@nbi.ku.dk (J.T.F.), aake@nbi.dk (\r{A}.N.)}\\ [\affilskip]
$^{10}$Department of Physics and Astronomy, University of Gent, Proeftuinstraat 86, Gent B-9000, Belgium; email: {\tt ameli@ulg.ac.be} \\ [\affilskip]
$^{11}$LUTH, Observatore de Paris-Meudon, 5 place Jules Jansen, Meudon Cedex 92195, France\\ email: {\tt helene.sol@obspm.fr} \\ [\affilskip]
$^{12}$Department of Physics and Astronomy, University of Alabama, Tuscaloosa, AL 35487, USA\\ email: {\tt pehardee@gmail.com} \\ [\affilskip]
$^{13}$Department of Physics and Astronomy, Clemson University, Clemson, SC 29634, USA\\ email: {\tt hdieter@g.clemson.edu}
}

\pubyear{2016}
\volume{324}  %% insert here IAU Symposium No.
\jname{New Frontiers in Black Hole Astrophysics}
\editors{A. Gomboc, ed.}
\begin{document}

\maketitle

\begin{abstract}
We study the interaction of relativistic jets with their environment, using 3-dimen-\\sional relativistic particle-in-cell simulations for two cases of jet composition: (i) electron-proton ($e^{-}-p^{+}$)  and (ii) electron-positron ($e^{\pm}$) plasmas containing helical magnetic fields. We have performed simulations of ``global" jets containing helical magnetic fields in order to examine how helical magnetic fields affect kinetic instabilities such as the Weibel instability, the kinetic Kelvin-Helmholtz instability and the Mushroom instability. We have found that these kinetic instabilities are suppressed and new types of instabilities can grow. For the $e^{-}-p^{+}$ jet, a recollimation-like instability occurs and jet electrons are strongly perturbed, whereas for the $e^{\pm}$ jet, a recollimation-like instability occurs at early times followed by kinetic instability and the general structure is similar to a simulation without a helical magnetic field. We plan to perform further simulations using much larger systems to confirm these new findings.

\keywords{plasmas, instabilities, magnetic fields, shock waves, relativity, galaxies: jets, active}
%% add here a maximum of 10 keywords, to be taken form the file <Keywords.txt>
\end{abstract}

\firstsection % if your document starts with a section,
              % remove some space above using this command.
\section{Introduction}
Particle-in-cell (PIC) simulations of collisionless shock formation and instability growth, such as the Weibel instability, the kinetic Kelvin-Helmholtz instability (kKHI), and the Mushroom instability (MI), have been performed to study magnetic field generation, particle acceleration, and emission of radiation with applications to astrophysical plasma jets (e.g., \cite[Alves et al. 2012]{alves12}, \cite[Nishikawa et al. 2003, 2005, 2013, 2014, 2016a,b]{nishikawa03,nishikawa05,nishikawa13,nishikawa14}). Furthermore, \cite{nishikawa16a} have performed global simulations involving injection of a cylindrical jet into an ambient plasma in order to investigate shock (Weibel instability) and velocity shear instabilities (kKHI and MI) simultaneously. Previously, these two processes have been investigated separately.

To our knowledge, \cite[Nishikawa et al. (2016b)]{nishikawa16b} were the first to present results of numerical PIC simulations of global relativistic jets containing helical magnetic fields. The kinetic features of secondary magnetic reconnection in a single flux rope undergoing internal kink instability are studied by means of three-dimensional particle-in-cell simulations where the single flux rope is modeled with a simple screw-pinch configuration as in  \cite{markidis14}. 

The presence of helical magnetic fields is suggested by twisted structures that have been observed in many active galactic nuclei (AGN) jets, from sub- to kiloparsec scales (e.g., \cite{lobanov01,perucho12,gomez16}). These jet structures were explained through relativistic magnetohydrodynamic (MHD) modeling, where simulations of current-driven (kink) instabilities were performed (e.g., \cite{mizuno15,singh16}). The main result obtained by \cite{nishikawa16b} has revealed that new types of shocks, similar to the recollimation shocks attained in relativistic MHD simulations, occur when a relativistic plasma jet contains a helical magnetic field. The simulations presented in this paper were designed to further test these new findings.

\section{Numerical methods and simulation setup}

We use a fully kinetic approach to model the formation of shocks in relativistic plasma jets containing helical magnetic fields. We apply PIC methods to numerically simulate the injection of a cylindrical relativistic jet with a Lorentz factor $\gamma = 15$  into an ambient plasma at rest, using a modified version (e.g., \cite{nishikawa03,nishikawa14,nishikawa16a}) of the TRISTAN code (\cite{buneman93}). 

For plasma composition, we use (i) an electron-proton ($e^{-}-p^{+}$) plasma with a realistic proton-electron mass ratio $(m_{\rm p}/m_{\rm e} = 1836)$ and (ii) an electron-positron ($e^{\pm}$) plasma. The simulations were performed with a numerical grid of $(L_{\rm x},L_{\rm y},L_{\rm z}) = (645\Delta,131\Delta,$ $131\Delta)$, where $\Delta=1$ is the cell size, and periodic boundary conditions in traverse directions. The plasma jet, with a radius of $r_{\rm jt} = 20\Delta$, is injected in the middle of the $y-z$ plane $((y_{\rm jc},z_{\rm jc}) = (63\Delta,63\Delta))$ at $x=100\Delta$. For the complete set of parameters, see \cite{nishikawa16b}.

The helical magnetic field structure is implemented using the equations in \cite{mizuno15}, with an exponential damping function for the magnetic fields external to the jet in order for plasma instabilities to grow (\cite{nishikawa16b}). However, our simulations use Cartesian coordinates. We set $\alpha =1$, thus eqs. (9), (10) and (11) from Mizuno et al. (2015) are reduced to eq. (\ref{eq.1}) and the magnetic field takes the form:
\begin{eqnarray}
B_{x} = \frac{B_{0}}{[1 + (r/a)^2]}, \,  \, \, \, \, \,  B_{\phi} =  \frac{(r/a)B_{0}}{[1 + (r/a)^2]}
\label{eq.1}
\end{eqnarray}

The toroidal magnetic field is created by a current $+J_{x}(y, z)$ in the positive $x$-direction, so that defined in Cartesian coordinates:
\begin{eqnarray}
B_{y}(y, z) =  \frac{((z-z_{\rm jc})/a)B_{0}}{[1 + (r/a)^2]}, \, \, \,\,  \,  \, 
B_{z}(y, z) =  -\frac{((y-y_{\rm jc})/a)B_{0}}{[1 + (r/a)^2]}.
\label{eq2}
\end{eqnarray}
Here $a$ is  the characteristic  length-scale of the helical magnetic field, $(y_{\rm jc},\, z_{\rm jc})$ is the jet center, and $r = \sqrt{(y-y_{\rm jc})^2+(z-z_{\rm jc})^2}$. The choice of helicity is defined by eq. (\ref{eq2}), and has left-hand polarity with positive $B_0$. 

In the simulations, the initial magnetic field amplitude parameter $B_{0}=0.1c$, $(c=1)$, $(\sigma = B^2/n_{\rm e}m_{\rm e}\gamma_{\rm jet}c^{2} =2.8\times 10^{-3})$, and $a = 5.0\Delta = 0.25*r_{\rm jt}$, with $r_{\rm jt} = 20\Delta$. 

\section{Simulation results}

Results of global jet PIC simulations containing helical magnetic fields were presented in \cite{nishikawa16b}. In the  $e^{-}-p^{+}$ jet case recollimation-like shocks are developed. In the $e^{\pm}$ jet case small recollimation structures are initially formed, and after instabilities have grown, currents extend outside the jet and the current density becomes turbulent. \cite{nishikawa16b} also compared their results to two different relativistic MHD simulations of (a) recollimation (\cite{mizuno15}) and (b) current-driven kink instability (\cite{singh16}).
\vspace{-0.00cm}
\begin{figure}[h]
\begin{minipage}[h!]{93mm} %95
\hspace{3.4cm} (a) 
\hspace{4.1cm} (b)  
\vspace*{-0.07cm}
\begin{center}
\includegraphics[scale=0.27]{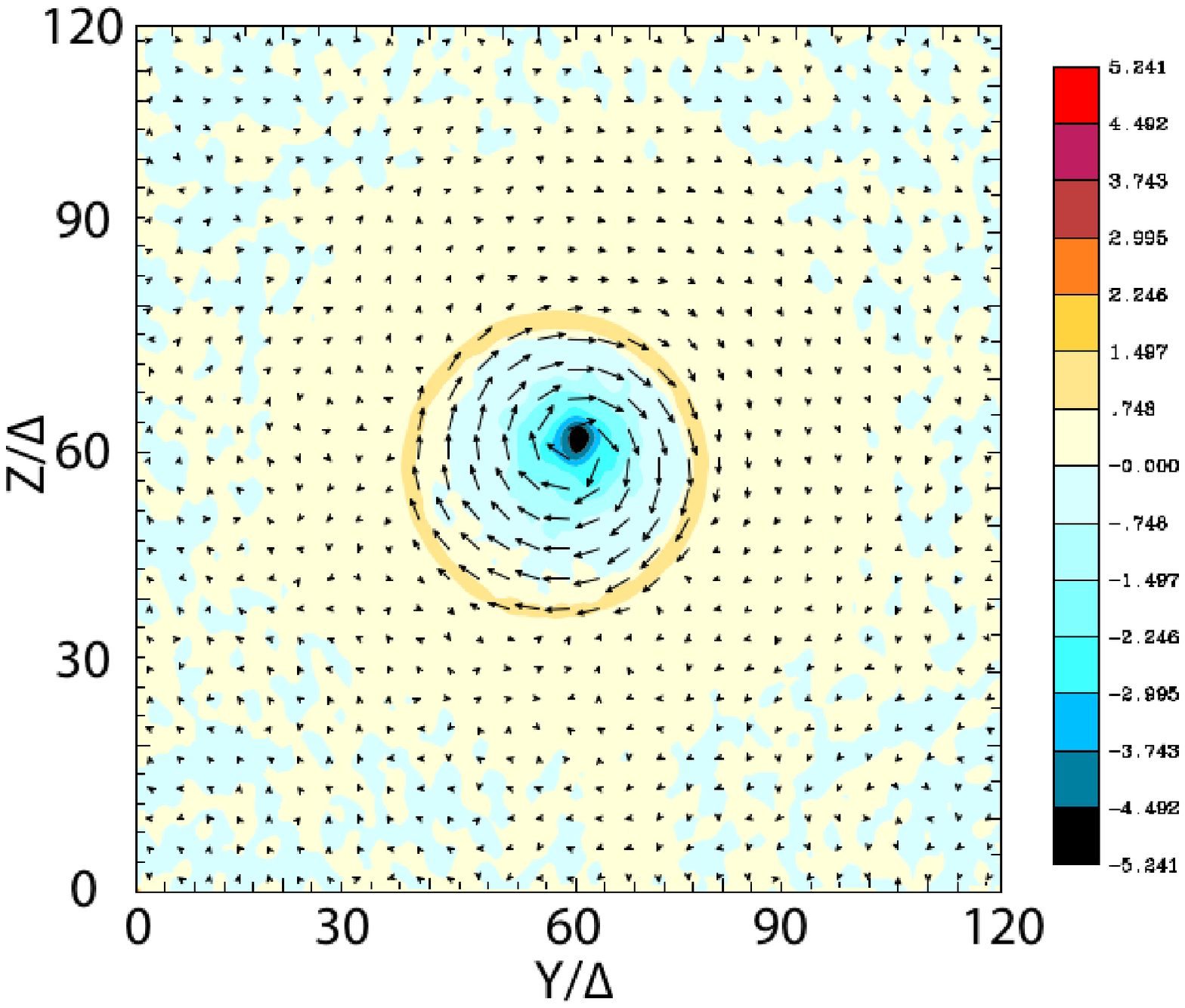}
\hspace*{-0.cm}
\includegraphics[scale=0.27]{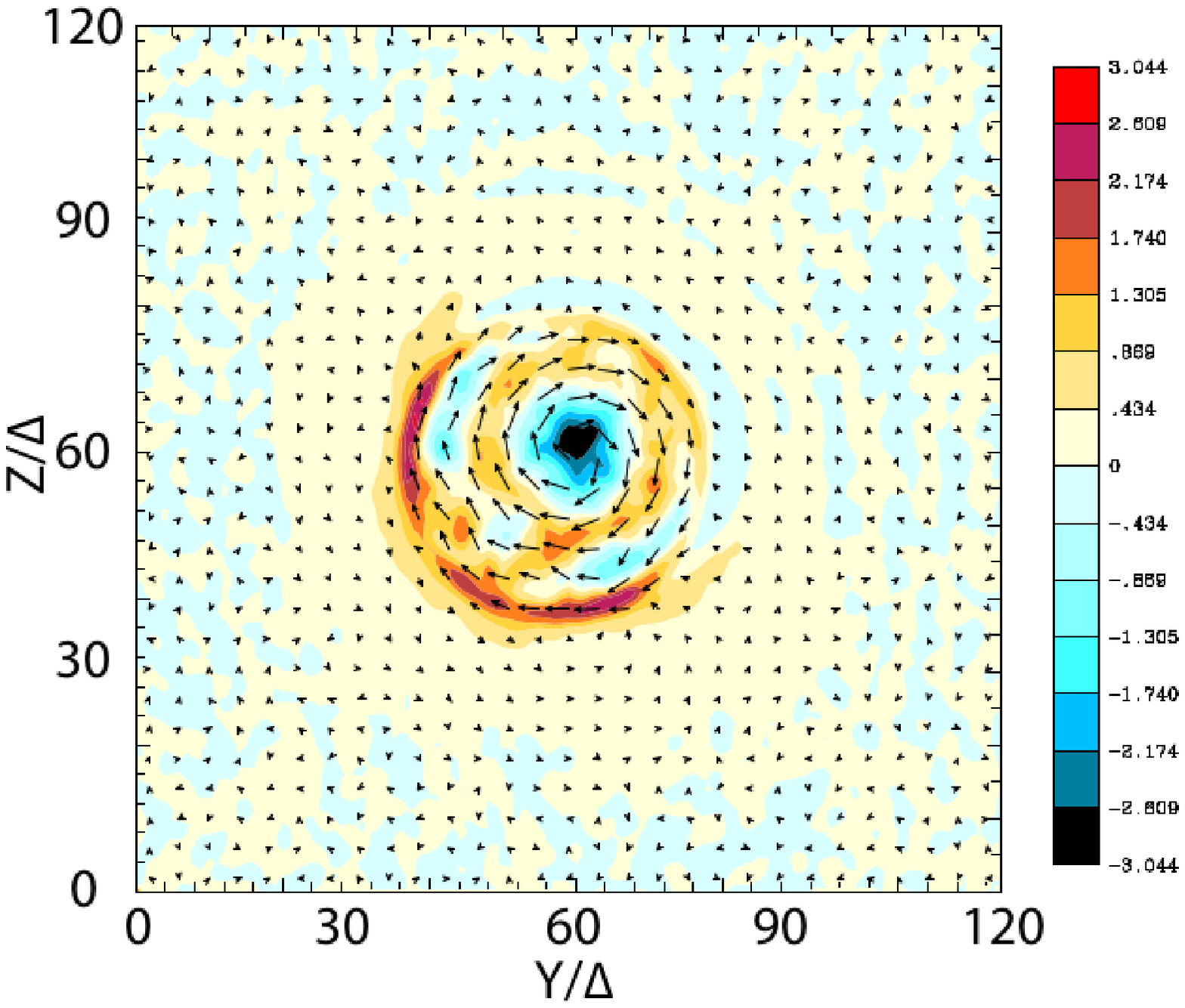}
\end{center}
\hspace{3.4cm} (c) 
\hspace{4.1cm} (d) 
\vspace*{-0.07cm}
\begin{center}
\includegraphics[scale=0.27]{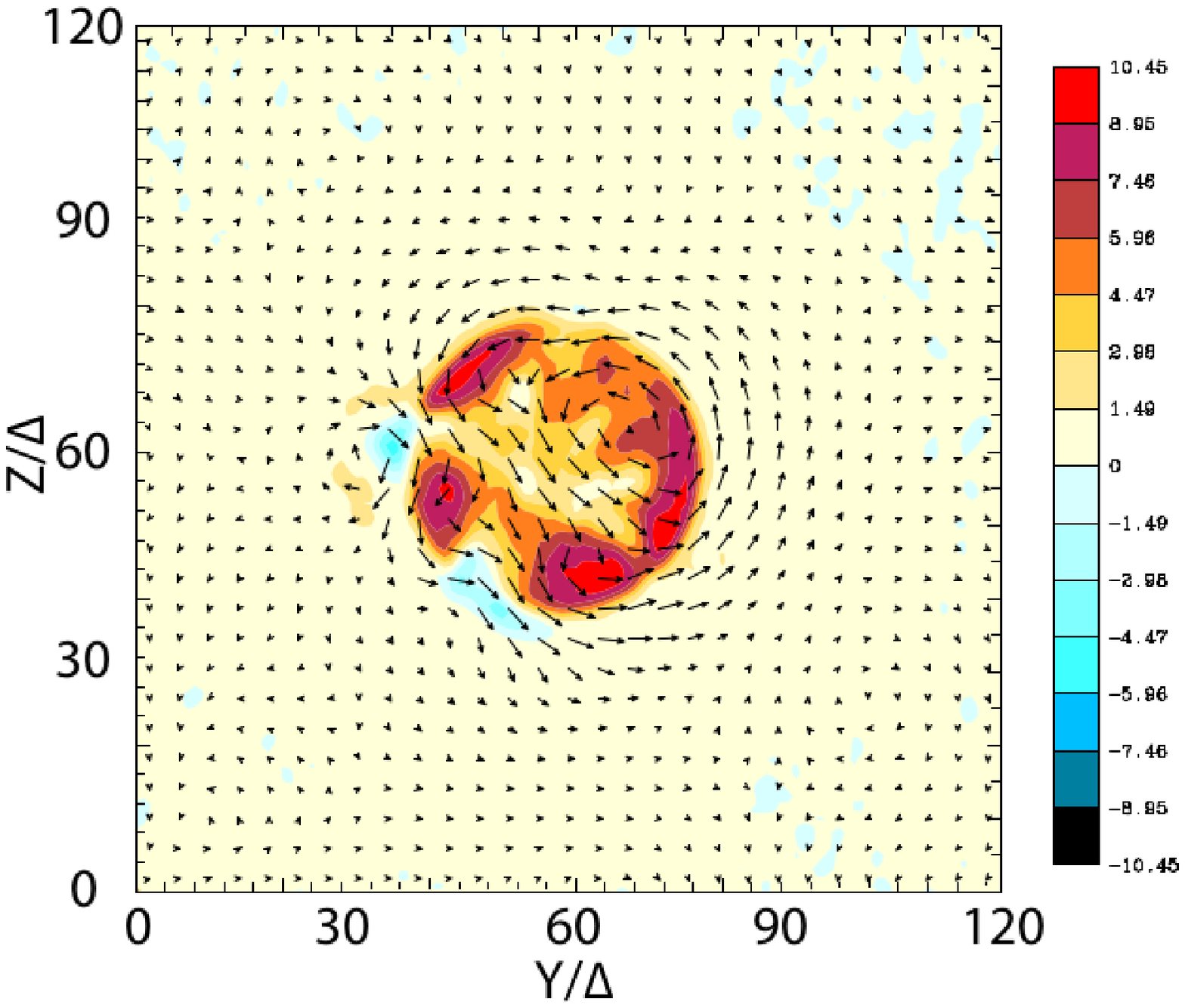}  %28
\hspace*{-0.cm}  
\includegraphics[scale=0.27]{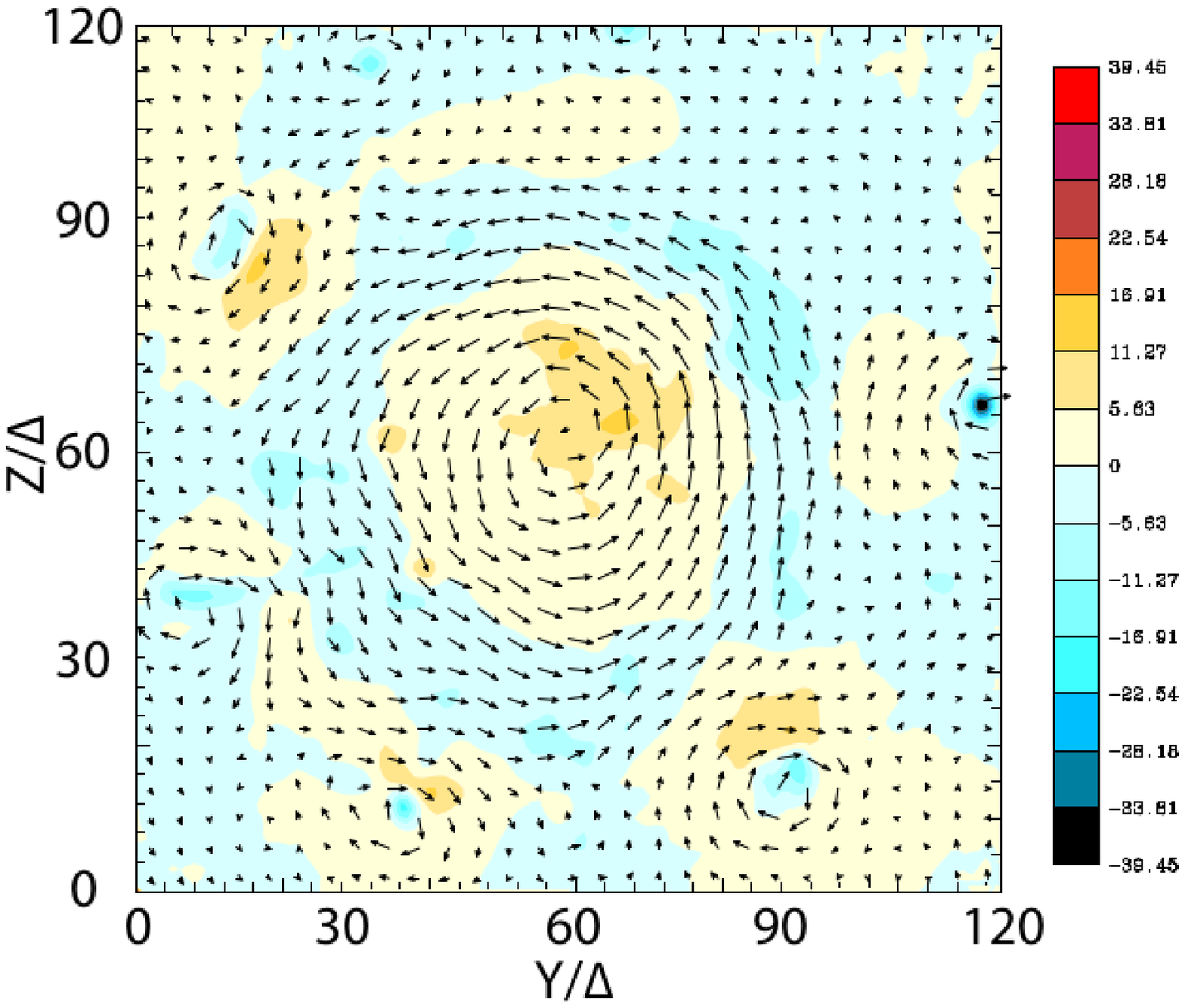}
\end{center}
\end{minipage}
\begin{minipage}[t]{35mm}
\vspace*{-3.7cm}
\caption{\footnotesize{Panels show 2D plots of the isocontour of $J_{x}$ in the $y-z$ plane: (a) and (c) for $e^{-}-p^{+}$ and (b) and (d) for $e^{\pm}$ plasma jets, at time 
$t=500\omega_{\rm pe}^{-1}$. Arrows show the magnetic fields ($B_{y,z}$). Panels (a) and (b) show $J_{x}$ at $x = 180 \Delta$, whereas panels (c) and (d) show $J_{x}$ at $x = 480 \Delta$. For both cases, the original left-handed polarity of $B_{y,z}$ found at $x = 180 \Delta$ has switched to right-handed polarity at $x = 480 \Delta$.}} 
\label{jxByz}
\end{minipage}
\end{figure}

In Figure \ref{jxByz} we show 2D isocontour plots of the $x$-component of the current density $J_{\rm x}$ in the $y-z$ plane for the electron-proton ($e^{-}-p^{+}$) case and the electron-positron ($e^{\pm}$) case at time $t=500\omega_{\rm pe}^{-1}$. We have found in both cases that, due to the growth of jet instabilities, the original left-handed (clockwise viewed from the jet front) polarity of the magnetic field ($B_{y,z}$) is switched to right-handed polarity, as shown in Fig. \ref{jxByz}.

At $x = 180 \Delta$ instabilities start to grow and, for the $e^{-}-p^{+}$ jet at $x = 480 \Delta$,  the magnetic field structure is distorted and $B_{y,z}$ shows linear polarity inside the jet, as the magnitude of $J_{x}$ decreases towards a null value.  This is what we might expect from recollimation-like shocks.

\section{Conclusions}

Using a fully kinetic approach, the simulation results presented in \cite{nishikawa16b} demonstrate the formation, at the microphysics level, of new types of shock structures in relativistic plasma jets due to the presence of helical magnetic fields, whereas kinetic instabilities, such as the Weibel instability, kKHI, and MI are suppressed. These new types of shocks have structures similar to those obtained using relativistic MHD methods (\cite{mizuno15,singh16}). Here, we have also shown that due to the presence of instabilities, the polarity of the magnetic field $B_{y,z}$ switches sign and, in the case of the $e^{-}-p^{+}$ jet, $B_{y,z}$ shows linear polarity inside the jet. 

Further systematic study of relativistic jets containing helical magnetic fields along with associated processes such as reconnection, turbulence, and generation of ultra-high energy particles will provide more advanced interpretation of observed phenomena such as gamma-ray burst (GRB) emission and polarized emission in Blazars and GRB jets.

\acknowledgments{This work is supported by NSF AST-0908010, AST-0908040, NASA-NNX09AD16G, NNX12AH06G, NNX13AP-21G, and NNX13AP14G grants. I.D. was partially supported by the NUCLEU Program (Contr. No. 4N/2016/PN16470201). The work of J.N. and O.K. has been supported by Narodowe Centrum Nauki through research project DEC-2013/10/ E/ST9/00662. Y.M. is supported by the ERC Synergy Grant ``BlackHoleCam - Imaging the Event Horizon of Black Holes''  (Grant No. 610058).  M.P.~acknowledges support through grant PO 1508/1-2 of the Deutsche Forschungsgemeinschaft. Simulations were performed using  Pleiades and Endeavor facilities at NASA Advanced Supercomputing (NAS), and using Gordon and Comet at The San Diego Supercomputer Center (SDSC), and Stampede at The Texas Advanced Computing Center, which are supported by the NSF. This research was started during the program ``Chirps, Mergers and Explosions: The Final Moments of Coalescing Compact Binaries'' at the Kavli Institute for Theoretical Physics, which is supported by the National Science Foundation under grant No. PHY05-51164. The first velocity shear results using an electron$-$positron plasma were obtained during the Summer Aspen workshop ``Astrophysical Mechanisms of Particle Acceleration and Escape from the Accelerators'' held at the Aspen Center for Physics (1--15 September 2013).}


\begin{thebibliography}{}
\bibitem[Alves, E. P. \etal\ (2012)]{alves12}
{Alves, E. P., Grismayer, T., Martin, S. F., Fi\'uza, F. et al.} 2012, \textit{ApJ} (Lett.), 746, L14

\bibitem[Buneman (1993)]{buneman93}
{Buneman, O.} 1993, in Computer Space Plasma Physics: Simulation Techniques and Software; Eds: Matsumoto, H., Omura, Y., Terra Scientific Publishing Company: Tokio, Japan, 1993, pp. 67-79 

\bibitem[G\'omez \etal\ (2016)]{gomez16}
{G\'omez, J. L., Lobanov, A. P., Bruni, G., Kovalev, Y. Y. et al.} 2016, \textit{ApJ}, 817, 2

\bibitem[Lobanov \& Zensus (2001)]{lobanov01}
{Lobanov, A. P. \& Zensus, J. A.} 2001, \textit{Science}, 294, 128

\bibitem[Markidis \etal\ (2014)]{markidis14}
{Markidis S., Lapenta, G., Delzanno, G. L., Henri, P. et al.} 2014, \textit{PPCF}, 56, 6, article id. 064010 

\bibitem[Mizuno \etal\ (2015)]{mizuno15}
{Mizuno, Y., G\'omez, J. L., Nishikawa, K.-I., Meli, A. et al.} 2015, \textit{ApJ}, 809, 38 

\bibitem[Nishikawa \etal\ (2016a)]{nishikawa16a}
{Nishikawa, K.-I., Frederiksen, J. T., Nordlund, \r{A}., Mizuno, Y. et al.} 2016a, \textit{ApJ}, 820, 94 

\bibitem[Nishikawa \etal\ (2016b)]{nishikawa16b}
{Nishikawa, K.-I., Mizuno, Y., Niemiec, J., Kobzar, O. et al.} 2016b, \textit{Galaxies}, 4, 38

\bibitem[Nishikawa \etal\ (2014)]{nishikawa14}
{Nishikawa, K.-I., Hardee, P. E., Du\c{t}an, I., Niemiec, J. et al.} 2014, \textit{ApJ}, 793, 60

\bibitem[Nishikawa \etal\ (2013)]{nishikawa13}
{Nishikawa, K.-I., Hardee, P. E., Zhang, B., Du\c{t}an, I. et al.} 2013, \textit{Ann. Geophys.}, 31, 1535

\bibitem[Nishikawa \etal\ (2005)]{nishikawa05}
{Nishikawa, K.-I., Hardee, P. E., Richardson, G., Preece, R. et al.} 2005, \textit{ApJ}, 622, 927

\bibitem[Nishikawa \etal\ (2003)]{nishikawa03}
{Nishikawa, K.-I., Hardee, P. E., Richardson, G., Preece, R. et al.} 2003, \textit{ApJ}, 595, 555

\bibitem[Perucho \etal\ (2012)]{perucho12}
{Perucho, M., Mart\'i-Vidal, I., Lobanov, A. P. \& Hardee, P. E.} 2012, \textit{A\&A}, 545, 65

\bibitem[Singh \etal\ (2016)]{singh16}
{Singh, C. B., Mizuno, Y. \& de Gouveia Dal Pino, E. M.} 2016, \textit{ApJ}, 824, 48
\end{thebibliography}
\end{document}